\documentclass[11p]{revtex4-2}

\usepackage{mathrsfs}
\usepackage{amsfonts}
\usepackage{amssymb}
\usepackage{amsmath}
\usepackage{graphicx}
\usepackage{color}
\usepackage{stackrel}

\newcommand{\sket}[1]{\vert #1 )}
\newcommand{\sbra}[1]{( #1 \vert}
\newcommand{\sbraket}[2]{( #1 \vert #2 )}
\newcommand{\be}{\begin{equation}}
\newcommand{\ee}{\end{equation}}
\newcommand{\bqa}{\begin{eqnarray}}
\newcommand{\eqa}{\end{eqnarray}}
\newcommand{\braket}[2]{\langle #1 \vert #2\rangle}
\newcommand{\ketbra}[2]{\vert #1 \rangle \langle #2 \vert}
\newcommand{\bra}[1]{\langle #1 \vert}
\newcommand{\ket}[1]{\vert #1 \rangle}
\newcommand{\expected}[1] {\langle #1 \rangle}

\begin{document}

\title{Real-time Krylov Diagonalisation for Open Quantum Systems}

\author{D. A. Herrera-Mart\'i}
\affiliation{Universit\'e Grenoble Alpes, CEA List, 38000 Grenoble, France}

\date{\today}

\begin{abstract}
In this chapter, we demonstrate how real-time quantum Krylov subspace methods can be adapted to investigate open quantum systems described by the Lindblad formalism. We apply these methods to a two-photon-driven superconducting Kerr resonator and illustrate their use in estimating the Liouvillian gap within the Kerr Cat qubit regime.\\

This is based on work done between November 2025 and January 2026 and on the talk given in early February 2026 at the Kwekfest 2026 conference, to celebrate L.C. Kwek's lifetime achievements [to be published by World Scientific].
\end{abstract}

\maketitle

\section{Introduction}

It is widely accepted that one of the earliest practical applications of quantum computing with economic value—beyond cryptography \cite{cain26}—will involve quantum-enhanced generation and analysis of classical datasets \cite{zhao26,chen24}. This could soon enable AI systems to design more efficient quantum hardware, a practice already established in classical inverse problems, shape optimization, and, more recently, chip design \cite{zheng23,khailany23}.

Since quantum technologies fundamentally rely on dissipative components, any quantum-enhanced method for the automated discovery of quantum hardware must inherently account for noisy, open quantum systems.  One could even argue that the first iteration of this loop has already taken place, since the introduction of \emph{Majorana 1 qubit} \cite{castelvecchi25}, whose underlying structure was for a long time the subject of quantum and classical simulation efforts \cite{alicea12,beenakker13}. Recently some work has explored how one might improve the design of quantum technologies using quantum computers \cite{naranjo26} .

Quantum technologies are becoming less overhead-intensive and quantum advantage appears closer than previously thought \cite{bravyi24, google25}.  At the same time, sub-exponential quantum speedups are generally considered insufficient for relevant applications for a series of reasons, including quantum clock speed, redundancy requirements and aymptotic constants \cite{hoeffler23}, and classical simulation techniques of quantum systems are pushing the boundary of quantum supremacy further deep into the realm of quantum fault-tolerance \cite{zhai26}. Consequently, future efforts are likely to concentrate on improving the Quantum Phase Estimation (QPE) primitive for different applications \cite{low19,su21},  wheras Variational Quantum Algorithms, despite their initial promise and heuristic speedups, might to become less relevant.

Recently, a new class of quantum algorithms known as Quantum Subspace Diagonalisation (QSD), designed for NISQ hardware, superseded variational algorithms as the state of the art in non-fault tolerant hardware \cite{cortes22,huggins20,klymko22, kirby23, motta24}. This is due, among others, to the fact that they do not require an external loop for ansatz optimisation, and that the subspaces these algorithms generate can be shown to be robust to certain kinds of errors, such as Trotter errors \cite{kirby24, epperly22}. Real-time QSD relies on having a quantum computer evolve an initial state $\ket{\psi_{0}}$ over several time intervals, i.e. implementing powers of the same time-evolution unitary operator. This yields the Krylov subspace $\textrm{span}\{ \ket{\psi_{0}},\ket{\psi_{1}}, \ket{\psi_{2}}\cdots, \ket{\psi_{D-1}}\}$, where $\ket{\psi_{k}} = \exp -i H k \tau \ket{\psi_{0}}$ are the time-evolved wavefunctions. Whereas classical Krylov methods (such as the Lanczos or Conjugate Gradient algorithms), implement reorthogonalisation on-the-fly while constructing the Krylov subspace, quantum circuits do not allow for this, and reorthonormalisation must be enforced via solving a Generalised Eigenvalue Problem (GEVP) $\bar H v = \lambda S v$ with:

\bqa
\bar H_{ij} &=& \bra{\psi_{i}}H\ket{\psi_{j}}\\
S_{ij} &=& \braket{\psi_{i}}{\psi_{j}}
\eqa

QSD marks a departure from the heuristic approach of variational algorithms by offering a more principled derivation of error bounds \cite{epperly22}, and has been experimentally demonstrated \cite{yoshioka25,yu25}. In contrast to Variational Quantum Algorithms, which are intrinsically biased by ansatz selection, the QSD methods are unbiased in principle, as with QPE, albeit with quadratic error scaling. For a given target accuracy $\epsilon$, the approximate cost of each algorithms is:

\bqa
C_{\textrm{QPE}} &\approx & O(1/poly(S))\left(O(poly(1/\epsilon)poly(N)) + C_{prep}\right)\\
C_{\textrm{QSD}} &\approx & O(1/poly(S)) (O(\frac{D^2(1+\eta)}{\epsilon^2\Delta^2}) + C_{prep}) + O(D^3) 
\eqa
with $S = |\braket{\psi_{GS}}{\psi_{0}}|$,  $\ket{\psi_{GS}}$ the ground state and $ C_{prep}$ is the initial state preparation cost \cite{lee23} (cost estimate for QSD is our own).  $\eta$ is a measure of gate noise, and the $1/\Delta^2$ dependence is a consequence of gap resolution via repeated measurements. Sampling error really is the bottleneck for these algorithms \cite{lee24} (for Toeplitz matrices the scaling is more benign in $D$ as only time differences are measured). The $O(D^3)$ cost comes from solving  of the generalised eigenvalue problem for dense matrices (see below). Both QPE and QSD rely on good initial overlaps for covergence, since they implement spectral filters of the initial state. Obtaining a good initial overlap be computationaly prohibitive for large strongly-correlated systems. Other sources of error, like ill-conditioning, are discussed below.

\section{Real-time QSD for Liouvillians}

Quantum Phase Estimation can be viewed as a spectral filtering procedure, where projective measurement realises a squared Dirichlet kernel \cite{nielsen10}. Its advantage stems from the fact that the central peak narrows exponentially with the number of ancilla qubits, yielding exponentially finer phase resolution. Similarly, quantum real-time Krylov diagonalisation can be understood as the optimsation of a (unitary) spectral filter, given fixed evolution times and initial states. To see this, consider that the overlap with the true eigenstates cannot be obtained by construction, so the proxy figure of merit to be minimised is the energy of the system given constraints on the vectors yielding a positive semidefinite overlap matrix. This involves minimising a Rayleigh quotient. If one defines $\ket{\Phi(w)} = \sum_i w_i \ket{\psi_{i}}$:

\be
R(w) = \frac{w^T\bar H w}{w^T S w},
\ee
minimising $R(w)$ under the constraint that the overlap matrix $S$ is positive semidefinite is equivalent to minimising the associated Lagrangian over the weights $w$:
\be
\mathcal{L}(w,\lambda) = w^T\bar H w - \lambda (w^T S w - K)
\ee
where K is a positive constant. Extremising the Lagrangian gives $\partial_w \mathcal{L}(w,\lambda) = 0 \rightarrow \bar H w = \lambda S w$. This means that, for given evolution time $\tau$, initial state $\ket{\psi_{0}}$ and Krylov dimension $D$, solving the GEVP gives the optimal ratio. Let $c_n = \braket{E_n}{\psi_0}$ and $f_w(E) =  \sum_k w_k \exp (-i E \tau k)$, then the Rayleigh quotient can be written as:

\be
R(w) = \frac{\sum_n |c_n|^2 E_n |f_w(E_n)|^2}{\sum_n |c_n|^2 |f_w(E_n)|^2}
\ee

Thefore, real-time QSD generates a spectral filter which is a sum of real amplitudes peaking at low energies. This interpretation reveals its connection to imaginary time evolution filtering, in which the filter is a real exponential.

It is further possible to optimise $\tau$,  $\ket{\psi_0}$ and $D$. The first obvious constraint on these parameters is that energy resolution is inversely proportional to the total time of the filter $T = D \tau$, i.e. $\delta E \sim 1/T$, so for the gap to be resolved one needs $T \geq 2\pi/\Delta$ and $\tau \geq 2\pi/(D\Delta)$. At the same time, the Nyquist criterion demands that the sampling time be smaller than the maximum energy to be resolved, i.e. $\tau \leq \pi/ E_{max}$, to avoid aliasing errors. Furthermore, $\tau$ should not be too small in order to avoid colinearities between different elements of the Krylov basis. Increasing $D$ seems like the only solution, but it leads to numerical instabilities. This can be seen by expressing the problem in the energy eigenbasis, in which the Vandermonde structure of the overlap matrix S becomes obvious \cite{kirby23}. Consider that $\exp{(-i H\tau k)} \ket{\psi_0} = V^\dag \Theta^k V \ket{\psi_0} = V^\dag \Theta^k \ket{\bar \psi_0}$, with $\Theta=\textrm{diag}(\theta_1,\theta_2,\cdots,\theta_N)$, $\theta_j = \exp{-iE_j \tau}$. Then, the Krylov matrix obtained by real-time unitary evolution is:

\be
K = V^\dag  \times \textrm{diag}( \braket{E_1}{\psi_0},\cdots \braket{E_N}{\psi_0}) \times \left[
\begin{array}{ccccc}
\theta^0_1 & \theta^1_1 & \cdots   & \theta^{(D-1)}_1 \\
\theta^0_2 & \theta^1_2 & \cdots   & \theta^{(D-1)}_2 \\
\vdots & \vdots & \ddots  & \vdots \\
\theta^0_N & \theta^1_N & \cdots  & \theta^{(D-1)}_N
\end{array}
\right]
\ee

This Krylov matrix has a Vandermonde structure and is therefore exponentially ill-conditioned on the linear dimensions, and so is the overlap matrix $S = K^\dag K$. This affects the numerical accuracy of the GEVP \cite{saad11}. The error in the (ground state) energy estimation is:

\be
|E_{GS} - \hat \lambda_0|\leq |E_{GS} - \lambda_0|_{\textrm{algorithmic}}  + |\lambda_0 - \hat\lambda_0|_{\textrm{numerical}}
\ee
where $\lambda_0$ is the ground state energy obtained from the GEVP (at infinite machine precision). The first part is dominated by gate and sampling noise, and is discussed extensively in the literature \cite{kirby24,lee24}. Here we focus on numerical error, since it needs to be kept as low as possible to make sure that the dominant source of error is the algorithmic one, and because it changes in a non-trivial way for Liouvillians. The numerical error is related to both the ill-conditioning of the overlap matrix and the norm of the Hamiltonian, $|\lambda_0 - \hat\lambda_0|_{\textrm{numerical}} \leq \varepsilon_{machine} \kappa(S) |H|$ . Roughly speaking, the GEVP will fail when the condition number $\kappa(S)$ reaches the same magnitude as the machine precision \cite{saad11}. In practice, thresholding strategies or regularisation are used to mitigate ill-conditioning of S due to both sampling noise and high D \cite{epperly22,kirby24,lee24}. The selection of $\tau$,  $\ket{\psi_{0}}$ and $D$ will be highly dependent on the problem. 

\subsection{Extension to Non-Hermitian, Non-Normal Operators}

We now consider a specific generalisation of QSD for Liouvillian superoperators, which have unitary and non-unitary dissipative  (which we assume in Lindblad form) components. In contrast to closed Hamiltonian systems, where $H=H^\dag$, the Liouvillian is generally non-Normal: $[L,L^\dag] \neq 0$, which implies the existence of left and right non-orthogonal eigenvectors, satisfying the bi-orthogonality condition, i.e. $\sbraket{\rho^{(L)}_i}{\rho^{(R)}_j} = \delta_{ij} \forall \sbra{\rho^{(L)}_i},\sket{\rho^{(R)}_j}$. Liouvillian eigenvalues are complex, symmetric and non-positive: 

\bqa
L \sket{\rho_k} &=& \lambda_k \sket{\rho_k} \\
\lambda_k &=& \pm i|\omega_k| - \gamma_k
\eqa
The steady state corresponds to the null eigenvalue, i.e. $L\sket{\rho_{SS}} = 0$ and the gap is defined as the real part of the second largest (as in closest to 0) eigenvalue: 

\be
\Delta = - \textrm{max  } \Re \lambda_k \equiv \gamma_{SM} 
\ee
and is associated with a slow mode of the Liouvillian dynamics. This slow mode corresponds to a long-lived, collective degree of freedom, which cannot be easily destroyed by local dissipative channels ($0 < \Re \lambda_k \ll 1$). In this case, the problem involves a maximisation rather than a minimisation, and it is not hard to verify that extremising the associated Lagrangian will still give a GEVP. 

The selection of the total evolution time $T$ and the initial state demand more substantial modifications. First of all, if one is interested in estimating slow-mode dynamics, it is necessary to resolve the gap, which might close rapidly in the system size \cite{minganti18}. This links up with the rationale mentioned in the introduction, i.e. being able to use a quantum computer to simulate slow modes of quantum systems that are hard to simulate classically.

\begin{figure}[t!]
  \includegraphics[width=\linewidth]{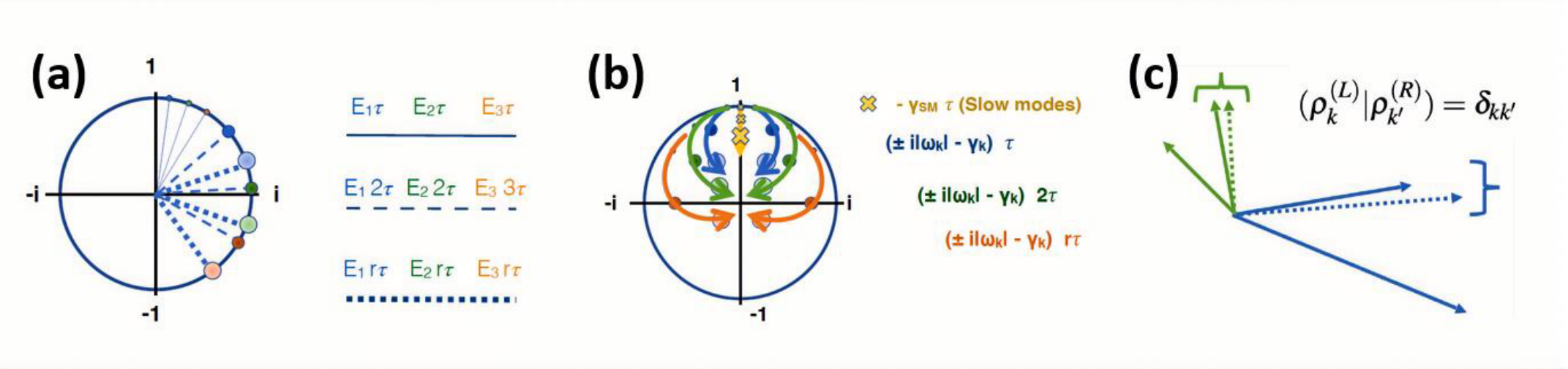}
  \caption{ {\bf(a)} Intuition from the complex circle: there is a tradeoff between frequency/energy resolution by
increasing $\tau$ and satisfying the Nyquist sampling criterion, to avoid time-evolved superkets to wind around the complex circle and cause artifacts like aliasing. {(\bf b)} Intuition from complex spirals: instead of clustering around the complex circle, the time-evolved (right) superkets spiral inwards. Fast decaying modes rapidly sink into the centre of the unit disk, and $\tau$ is not constrained by the Nyquist's criterion. The slow modes tend to survive for longer and they are captured in the Krylov representation {(\bf c)} Pictorial representation of the effect of small perturbations on the Liouvillian eigenstates. In order to ensure exact biorthogonality, the reoriented eigenstates must increase in norm.} 
\label{fig:circles} 
\end{figure}

In the Hamiltonian case, eigenstates are real and time-evolution is oscillatory, which implies the Nyquist sampling bounds to avoid artifacts (see Fig.\ref{fig:circles}(a)). In the Liouvillian case, if one focuses on resolving the gap, it is possible to extend the total sampling time by increasing $\tau$, instead of D (which as we have seen, can severely affect numerical stability). The underlying reason is that the steady state of the Liouvillian is an attractor for all states:

\be
\sket{\rho(t)} = \sket{\rho_{SS}} + \sum_k c_k(0) e^{i \omega_k t - \gamma_k t} \sket{\rho_k}
\ee

By construction, the contribution to the trace only comes from the steady state. Therefore, long evolution times will result in the loss of all signal except for slowly decaying modes, i.e. $\sket{\rho(t\gg1)} \approx \sket{\rho_{SS}} + c_{SM}(0) \exp(i \omega_{SM} t - \gamma_{SM} t)\sket{\rho_k}$ (see Fig.\ref{fig:circles}(b)). The Krylov subspace obtained this way is:
\be
\textrm{span}\{ \sket{\rho_{0}}, e^{L\tau}\sket{\rho_{0}}, e^{L2\tau}\sket{\rho_{0}}\cdots, e^{L(D-1)\tau}\sket{\rho_{0}}\}
\ee

\begin{figure}[h!]
  \includegraphics[width=\linewidth]{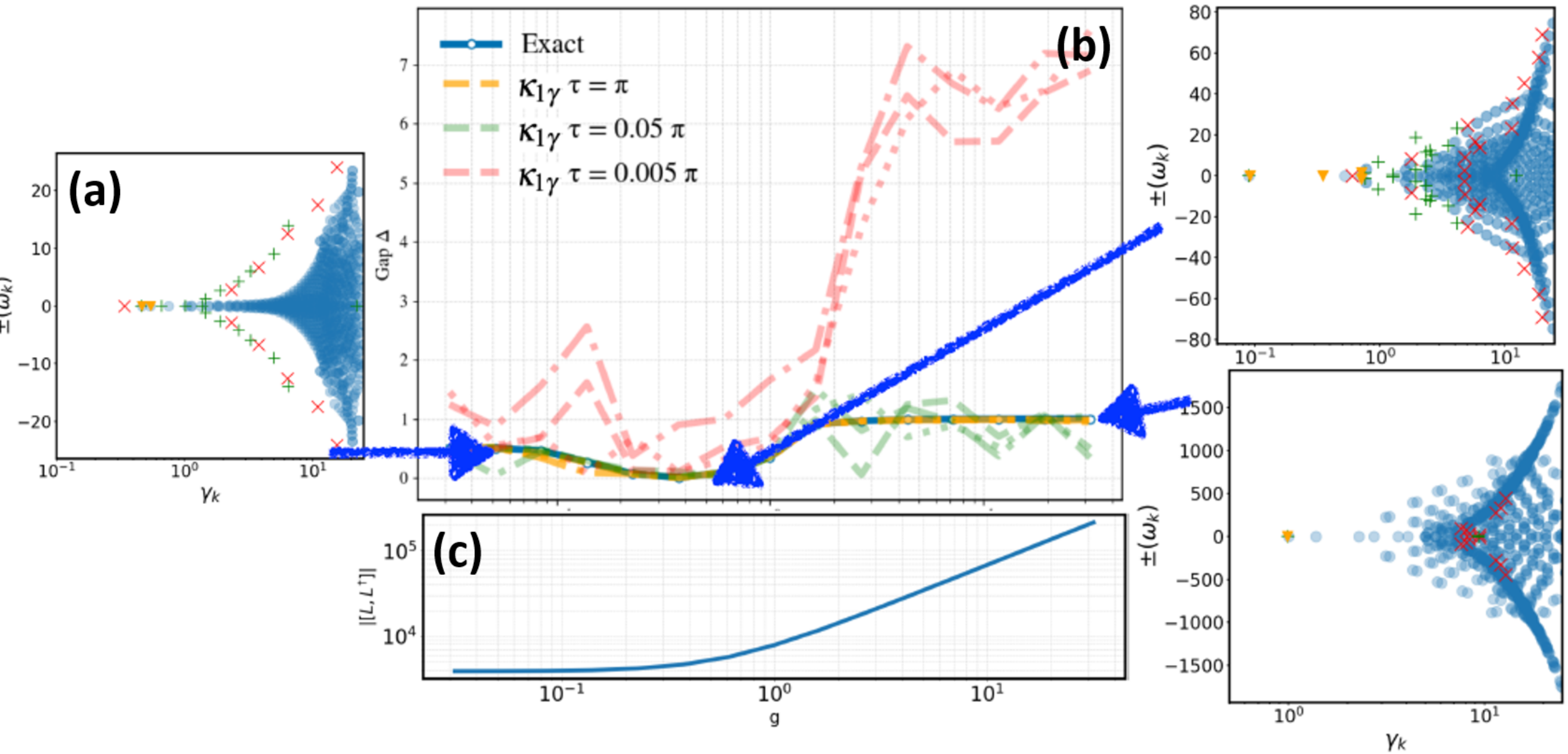}
  \caption{{\bf(a)} The Liouvillian spectra and associated reconstructions in different regimes (color code: orange $\bigtriangledown$, green $+$, red $\times$). {\bf(b)} Different gap reconstructions for increasing evolution time $\tau$. Longer $\tau$ results in better estimates, and poor ones seem to degrade as non-normality increases. Each experiment is repeated three times. {\bf(c)} Non-normality measured by the commutator of the Liouvillian and its adjoint.} 
\label{fig:long} 
\end{figure}

Elements of this basis are superkets of decreasing norm, as a consequence of damping modes of the Liouvillian. The filter interpretation is that, for long times, modes with non-zero oscillatory components (those typically in the ``wings" of the complex plane) tend to be washed out via phase cancelation and only slow modes survive (see Fig.\ref{fig:long}, in which only a handful of points are recovered from the total spectrum). 

Another important difference from the Hermitian QSD is that, in the dissipative case, the state corresponding to the largest (i.e. least negative) eigenvalue is the steady state, so the trace-1 sector does not carry signal useful for real-time QSD. In fact only the trace-0 sector of the initial state will give rise to a signal in the Krylov subspace (see Apendix I), which allows us to estimate the slow model eigenvalue $\lambda_{SM}$. In this case, the GEVP becomes significantly more ill-conditioned as a result of non-orthogonality of the Liouvillian eigenbasis. Non-normality entails high sensitivity to perturbations, as synthetised in the Bauer-Ficke bound \cite{saad11}. If V diagonalises L, i.e $L = V^{-1} diag(\vec \lambda) V$, them the numerical error of the GEVP for non-Normal operators can be shown to be:
\be
\epsilon_{\textrm{numerical}} \propto  \varepsilon_{\textrm{machine}}\kappa(V) \kappa(S) |L|
\ee
which can intuitively be undestood as arising from the biorthgonality requirement of non-Normal operators. Given that the direction of one right (left) eigenvector conditions the direction of all left (right) eigenvectors, a small perturbation in one vector has effects on all the subspace at once. If a right eigenvector $\sket{\rho^{(R)}_j}$ is perturbed in such a way that it forces a left eigenvector $\sbra{\rho^{(L)}_k}$ to change direction to ensure bi-orthogonality, the only way the corresponding right eigenvector $\sket{\rho^{(R)}_k}$ can continue to satisfy bi-orthogonality is to grow in norm (see Fig.\ref{fig:circles}(c)). This means that small fluctuations in the matrix entries can result in large increases in norm of the perturbed eigenvectors. This entails high sensitivity to noise.

Althought significant work has already been devoted to combining Krylov methods and Liouvillian superoperators, either for operator growth \cite{bhattacharya22,liu23,bhattacharjee24,bhattacharyya25}, or to reduce the overhead of unitary methods \cite{byrne24}, as far as we are aware, none of these studies focuses on real-time QSD to study dissipative dynamics.

\section{A Simple Kerr Cat Qubit}

We will now demonstrate this method for a simple dissipative system which is relevant for building robust qubits: the Kerr Cat qubit (KCQ) \cite{mirrahimi14,venkatraman24,grimm20,ding25}. This is one of the two approches in the design of protected Cat qubits, the other one being about  reservoir engineering so that dominant noise is a symmetry of the qubit \cite{mirrahimi14,leghtas15,lescanne20} and are promising architectures for quantum computing \cite{sivak23}. Admittedly, both single KQCs and their engineered-bath cousins are not difficult to simulate clasically. However, lattices of Cat qubits can quickly become computationally expensive in some regimes \cite{rota19}. 

In superconducting circuit architectures, Kerr-like interactions can be engineered using a SNAIL circuit (an asymmetric Josephson junction loop composed of one small junction and three larger junctions). Under magnetic flux bias, the circuit exhibits tunable cubic and quartic (Kerr-type) nonlinearities \cite{venkatraman24,grimm20} (this means that an appropriate rotating frame has to be defined, but we will not dwell in these details). The system is squeezed by a two-photon pump, so that the resulting (simplified) Hamiltonian is:

\be
H_{KCQ} = \delta a^\dag a - K a^{\dag 2}a^{2} + g (a^{\dag 2} +a^{2})
\ee
where $\delta$ is the detuning between the pump and the bare system, $K$ is the Kerr non-linearity term and $g$ is the two-photon pump strength. In the case of zero detuning, the factorisation $H_{KCQ} = -K(a^{\dag 2} - g/K)(a^{2} - g/K) +
 g^2 / K$ shows that the energy is minimised for two equivalent coherent states $\ket{\alpha = \pm \sqrt{g/K}}$. This Hamiltonian is symmetric under $a \leftrightarrow -a$, i.e there is a parity operator $\Pi = \exp(-i\pi a^\dag a) = (-1)^{\hat n}$ such that $[\Pi, H_{KCQ}] = 0$. This implies a ground state degeneracy, and its true eigenvectors are the symmetric and antisymmetric Cat states $\ket{\mathcal{C}^\pm} = \mathcal{N}^\pm (\ket{+\alpha} \pm \ket{-\alpha})\equiv\ket{\pm}$ (with normalisation $\mathcal{N}^\pm$).  $\ket{\mathcal{C}^+}$($\ket{\mathcal{C}^-}$) consist of even (odd) number of photons and are the true ground states of $H_{KCQ}$. The coherent states $\ket{+\alpha}\equiv\ket{0}$ and $\ket{-\alpha}\equiv\ket{1}$ break this symmetry.

The logical operators are $\bar Z = a/\alpha$ and $\bar X = \Pi$:

\bqa
\bar Z \ket{\pm \alpha}  &=& \pm \ket{\pm \alpha} \\
\bar X \ket{\pm \alpha} &=& \ket{\mp \alpha}
\eqa
The definition for $\bar X$ follows from applying the parity operator to photon number states:

\be
\Pi \ket{+ \alpha} \propto \sum^\infty_n \frac{\alpha^n}{\sqrt{n!}}e^{-i\pi a^\dag a} \ket{n} = \sum^\infty_n \frac{(-\alpha)^n}{\sqrt{n!}} \ket{n} \propto \ket{- \alpha}
\ee

It is easy to verify the anticommutation relations of the qubit operators $\{\bar X, \bar Z \} \equiv \{\Pi, a\} = 0$ in the number basis:

\be
(\Pi a + a \Pi)\ket{n} = \left((-1)^{n-1}\sqrt{n} + \sqrt{n} (-1)^{n}\right)\ket{n-1} = 0
\ee

\begin{figure}[h!]
  \includegraphics[width=\linewidth]{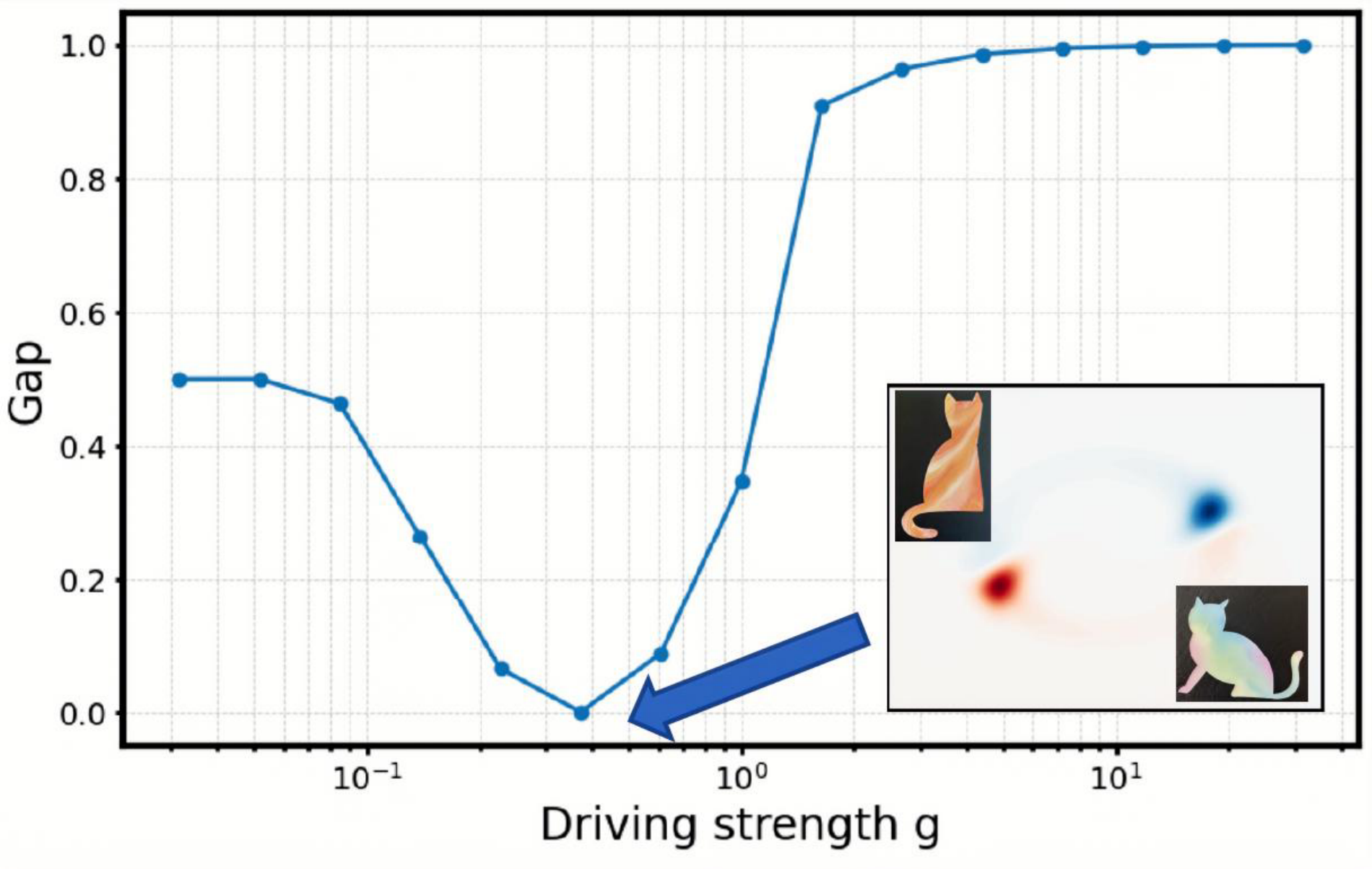}
  \caption{The Kerr Cat Regime coincides with the onset of the Liouvillian gap closing. This is because a slow mode, corresponding to the coherences and population imbalances, becomes long-lived. The gap is reduced as $\propto \exp -\alpha^2$, and corresponds to the phase-flip rate. At very high drive, the symmetry is restored and the gap reopens. Inset: Wigner representation of the slow mode $\sket{\rho_{SM}}$.} 
\label{fig:cat} 
\end{figure}

On top of this, we will consider a simplified Liouvilllian, accounting only for single photon loss: 

\be
L_{KCQ} \rho =  - i [H_{KCQ}, \rho] + \kappa_{1\gamma} \mathcal{D}[a]\rho
\ee
where $\mathcal{D}[a] \rho = a \rho a^\dag - \frac{1}{2}\{a^\dag a, \rho\}$ is the Lindbladian dissipator. A more detailed understanding of ground state degeneracy comes from the Heisenberg equations of motion (see Appendix II), which describe the emergence of a biased-noise qubit regime as the onset of a phase transition. Indeed, at roughly $g \geq \kappa_{1\gamma}/4$ (cf. Fig.\ref{fig:cat}), the photon number in the cavity increases so that the relative strength of the Kerr non-linearity becomes comparable to that of the drive, and stabilises a double well potential. 

Photon loss induces parity flip noise of the Cat states and stabilises the coherent states $\ket{\pm\alpha}$. Photon dephasing leads to $\bar X$ errors, and can come either from systematic coherent sources (of type $R(\beta) = \exp(-i \beta a^\dagger a)$, which can be calibrated) or from purely high frequency noise, the amplitude of which can be estimated via the WKB approximation \cite{venkatraman24}. This entails a bit-flip error exponentially reduced in $g$, and a polynomially increase in phase-flip errors \cite{mirrahimi14}.

\begin{figure}[h!]
  \includegraphics[width=\linewidth]{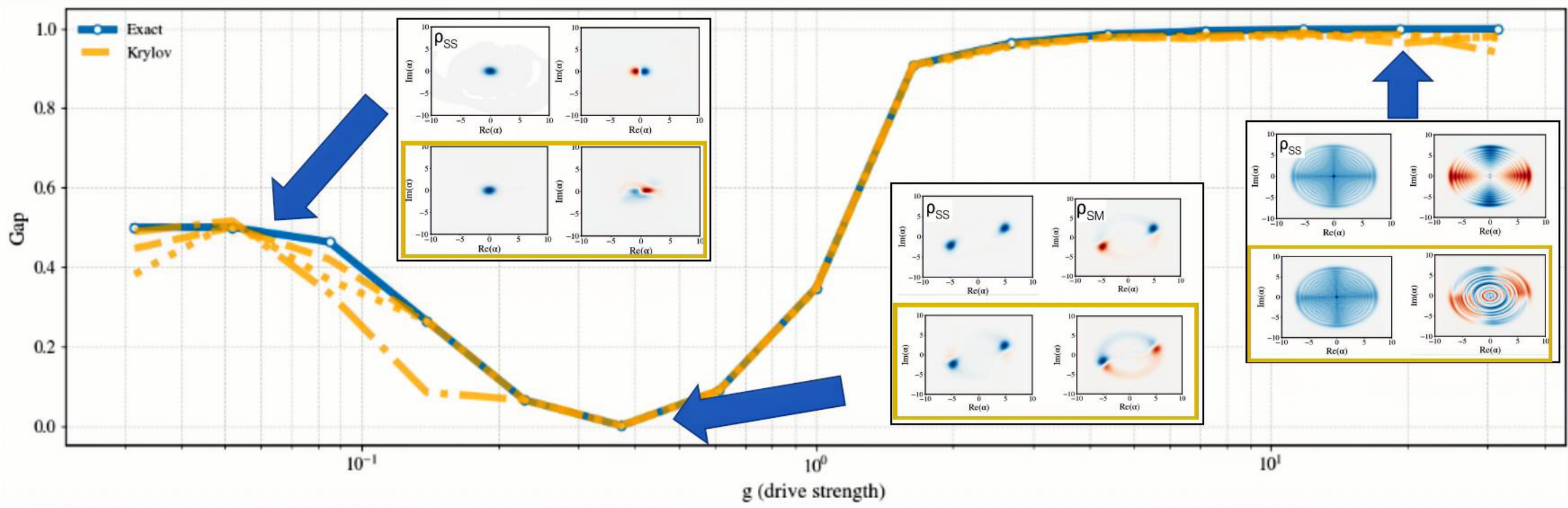}
  \caption{Long times allow for a faithful reconstruction of the gap across several orders of magnitude. In this simulation we have access to the ($900\times 20$) Krylov matrix which can be used to reconstruct the Wigner representation of the Liouvillian eigenstates. Insets: Real and reconstructed Liouvillian eigenstates $\sket{\rho_{SS}}$ and $\sket{\rho_{1}}$(left: damped regime, center: in the biased-noise Cat regime the slow mode features two lobes, right: strong drive, squeezed regime). In the very strong field limit, the dominant term is the two-photon pump. And the gap opens again.
} 
\label{fig:long2} 
\end{figure}

The steady state of $L_{KCQ}$ is $\rho_{SS} \propto \ketbra{+\alpha}{+\alpha} + \ketbra{-\alpha}{-\alpha}$ which, in the language of computation, is the completely mixed qubit. It is the fixed point towards which the photon-loss dynamics converge in the long-time limit. The first eigenstate (associated to the first non-zero eigenvalue) of $L_{KCQ}$ is $\rho_{1} \propto \ketbra{+\alpha}{+\alpha} - \ketbra{-\alpha}{-\alpha}$. The Cat qubit is protected in the sense that information becomes non-local (i.e. local dissipative dynamics cannot destroy it). The closing of the Liouvillian gap is associated to the emergence of a protected computational subspace spanned by $\rho_{SS}$ and $\rho_1$ (i.e. steady state and slow mode), quasi-stabilised by dissipative dynamics, and is related to the theory of dissipative phase transitions \cite{minganti18,zhang21,beaulieu25} (in our finite size case, the gap remains open but is reduced).

Our goal is to apply the theory of real-time QSD developed in previous section to estimate the gap for a sweep of diferent field strengths $g$. Our simulations were done for $\kappa_{1\gamma}=1$, $\delta=\kappa_{1\gamma}/5$, $K=\kappa_{1\gamma}/20$, a maximum of 30 photons and $D=20$. This corresponds to a Krylov matrix of dimension $900\times 20$ and a compression factor of $\sim 98\%$.
As explained previously, it is possible for the evolution time in QSD to go beyond the Nyquist sampling criterion because of dissipative dynamics, since fast modes will shrink faster towards the centre of the unit complex disk (see Fig.\ref{fig:circles}(b) ) and will not result in artifacts that wind around the complex circle. This allows to reconstruct the gap for a smaller amount of Krylov dimensions and limit the ill-conditioning of the overlap matrix.

Long evolution times $\tau$ lead to good resolution of the gap, provided that they are not too large (as it would erase the slow mode). We computed the non-normality of the Liouvillian as $|[L_{KCQ}^\dag(g), L_{KCQ}(g)]| / |L_{KCQ}(g)|$ and verified that the gap becomes more difficult to estimate for highly non-Normal Liouivillians.

\section{Conclusions and Acknowledgements}

This chapter hopefully demonstrates that it is possible to use real-time QSD to study open quantum systems (in Lindblad form). These methods must be substantially modified in order to account for the structural differences between Hamiltonian and Liouvillian systems. Here they were applied to a simplified version of a noise-biased protected qubit: the Kerr Cat qubit.

I acknowledge stimulating conversations with Clemens Gneiting and Neill Lambert at RIKEN. This work is supported by the France 2030 HQI project (ref. ANR-22-PNCQ-0002). Special thanks to S. and C. Herrera-Castrioto for their depictions of the cats in Fig.\ref{fig:cat}.

\begin{flushright}
Happy Birthday Kwek!
\end{flushright}  

\bibliographystyle{plain}

\pagebreak
\section*{Appendix I: Circuit Simulation for Lindbladian Dynamics}

Dissipative dynamics in Lindblad form can be implemented on a quantum circuit provided that we (1) double the size of the quantum register to encode flattened density operators, and (2) perform a moderate amount of classical post-processing \cite{kamakari22}.  The non-unitary part is implemented by Quantum Imaginary Time Evolution (QITE) algorithm, so the superket remains normalised throughout evolution, i.e. $\sbraket{\rho(t)}{\rho(t)} = 1$ for all times. This is in contrast with the trace constraint enforced by Liouvillian evolution, i.e. $\sbraket{I}{\rho (t)} = 1$. Ref. \cite{kamakari22} shows how this can be accounted for by performing frequent estimations of the trace.

\begin{figure}[h!]
  \includegraphics[width=\linewidth]{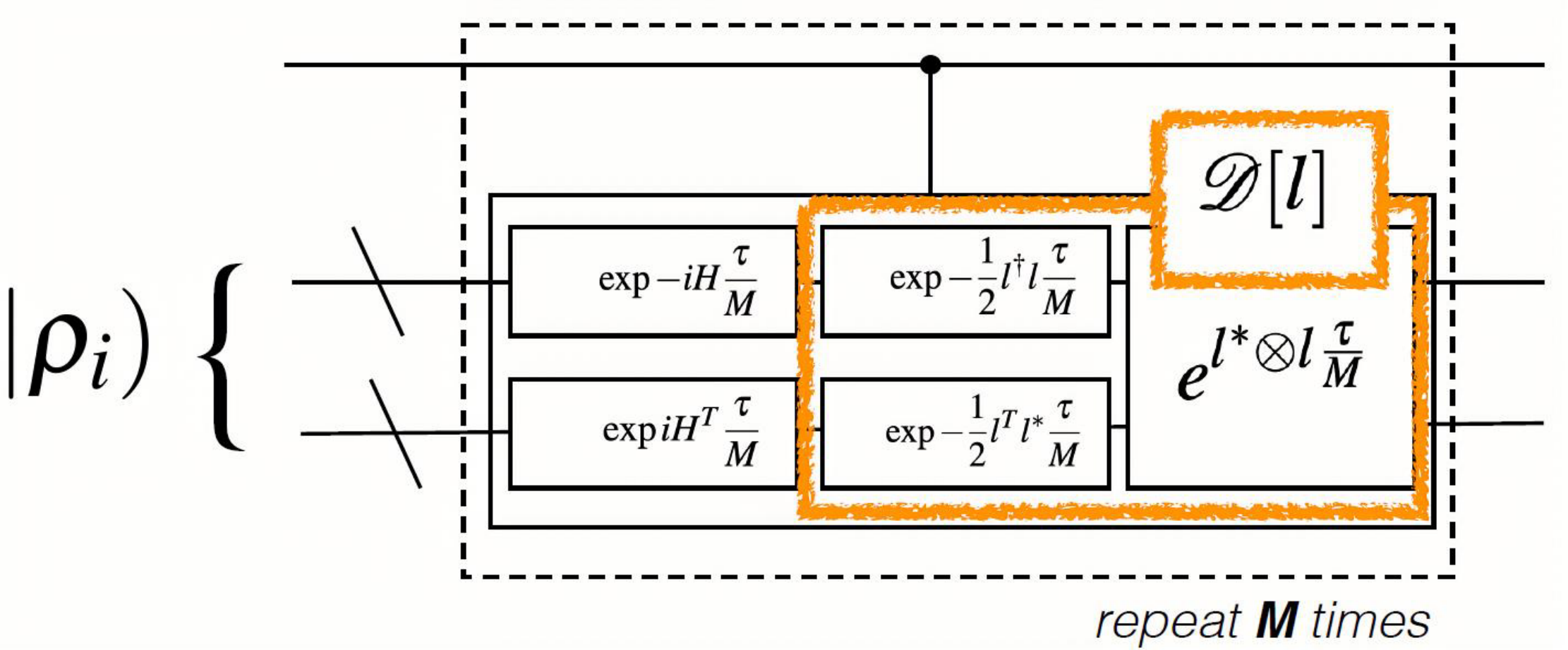}
  \caption{To simulate dissipative dynamics with a quantum computer, one can Trotterise reversible unitary circuits with QITE for the Lindbladian dissipator $\mathcal{D}[l]$:(adapted from ref. \cite{kamakari22}). Caveats: (i) the number of qubits doubles, (ii) long-time evolution is needed while keeping error low: deep circuits are needed, (iii)  non-unitary evolution is implemented with QITE, so the superkets remain normalised, and (iv) initial states must be valid density operators of trace 1.} 
\label{fig:qcircuit} 
\end{figure}

The resulting circuit alternates unitary and non-unitary dynamics. Crucially, the Trotterisation between unitary and non-unitary terms implies that the circuit needs to be fairly deep, which makes this approach only feasible together with some amount of error correction. The initial superkets must be valid density operators: 
\be
\sket{\rho(0)} = \frac{1}{N}\sket{I} + \sket{\rho_i}
\ee
where $\sket{\rho_i}$ amounts to initial coherences or population imbalances. For the initial state in the simulations of the KCQ, we uniformly sampled and subtracted random combinations of Kerr Hamiltonian eigenstates:
\be
\sket{\rho_i} = \textrm{vect}\left(\sum_{k,k'} c_{k,k'} (\ketbra{\psi_k}{\psi_k} - \ketbra{\psi_{k'}}{\psi_{k'}})\right)
\ee

Other simulation approaches for non-unitary quantum evolution have been tailored to operate on near-term devices \cite{li25}.

\section*{Appendix II: Symmetry Breaking in Kerr Cat qubits}

Cat qubits can be understood as emerging from symmetry breaking of a driven Kerr resonator by looking at the Heisenberg equations of motion. They can obtained by solving $\dot A = L_{KCQ}^\dag A$:

\bqa
\frac{d}{dt}\expected{a} = i\left(\delta \expected{a} + 2 K \expected{a^\dag a a} - 2 g \expected{a^\dag}\right) - \frac{\kappa_{1\gamma}}{2} \expected{a}\\
\frac{d}{dt}\alpha =i\left(\delta \alpha + 2 K |\alpha|^2\alpha - 2 g \alpha^*\right) - \frac{\kappa_{1\gamma}}{2} \alpha\\
\frac{d}{dt}\alpha = 0 \textrm{   } (\delta, \kappa_{1\gamma}=0) \rightarrow |\alpha|^2 = g / K
\eqa
where in the first line we assume normal ordering, and in the second line we have applied the mean field approximation ($\expected{a^\dag a a}\approx\expected{a^\dag}\expected{a}\expected{a}$) and  $\expected{a}=\alpha$. These equations highlight the connection with the Landau theory of continuous phase transitions: if one extremises the functional $F(\alpha) = K/4 |\alpha|^4 - g/2 |\alpha|^2$, one gets two minima at $\alpha = \sqrt{g/K}$. The presence of detuning renormalises slightly the steady state.

We can see that if the two-photon driving field is above roughly $\kappa_{1\gamma}/4$, and tuning on $K$ leads to a non-zero expected value for $\expected{a}$. One can see how, above a critical driving strength, the photon number suddenly increases, leading to two coherent states specularly opposed. This is a hallmark of three aspects of the same phenomenon: (i) non-locality of the ground state and protection from local dissipative channels, (ii) a (quasi-)degeneracy of the low-energy subspace of the Hamiltonian allowing for symmetry breaking and (iii) the closing of the Liouvillian gap and the emergence of a slow mode. The magnitude $\expected{a} = \pm \alpha$ is therefore the quantum-optics equivalent of the magnetisation in simple lattice systems. This, again, sheds light on the interpretation of photon loss and dephasing as logical operators.

\end{document}